# Enhanced creep performance in a polycrystalline superalloy driven by atomic-scale phase transformation along planar faults

Lola Lilensten[1,#*], Stoichko Antonov[1], Baptiste Gault[1,2], Sammy Tin[3], Paraskevas Kontis[1*]

[1] Max-Planck-Institut für Eisenforschung, Max-Planck Straße 1, 40237 Düsseldorf, Germany

[2] Department of Materials, Imperial College, South Kensington, London SW7 2AZ, UK

[3] Illinois Institute of Technology, 10 W. 32nd Street, Chicago, IL, 60616, USA

[#] now at Chimie ParisTech, PSL University, CNRS, Institut de Recherche de Chimie Paris, 75005 Paris, France.

* corresponding authors: Lola.lilensten@chimieparistech.psl.eu, p.kontis@mpie.de

**Abstract**

Predicting the mechanical failure of parts in service requires understanding their deformation behavior, and associated dynamic microstructural evolution up to the near-atomic scale. Solutes are known to interact with defects generated by plastic deformation, thereby affecting their displacement throughout the microstructure and hence the material's mechanical response to solicitation. This effect is studied here in a polycrystalline Ni-based superalloy with two different Nb contents that lead to a significant change in their creep lifetime. Creep testing at 750°C and 600 MPa shows that the high-Nb alloy performs better in terms of creep strain rate. Considering the similar initial microstructures, the difference in mechanical behavior is attributed to a phase transformation that occurs along planar faults, controlled by the different types of stacking faults and alloy composition. Electron channeling contrast imaging reveals the presence of stacking faults in both alloys. Microtwinning is observed only in the low-Nb alloy, rationalizing in part the higher creep strain rate. In the high-Nb alloy, atom probe tomography evidences two different types of stacking faults based on their partitioning behavior. Superlattice intrinsic stacking faults were found enriched in Nb, Co, Cr and Mo while only Nb and Co was segregated at superlattice extrinsic stacking faults. Based on their composition, a local phase transformation occurring along the faults is suggested, resulting in slower creep strain rate in the high-Nb alloy. In comparison, mainly superlattice intrinsic stacking faults enriched in Co, Cr, Nb and Mo were found in the low-Nb alloy. Following the results presented here, and those available in the literature, an atomic-scale driven alloy design approach that controls and promotes local phase transformation along planar faults at 750°C is proposed, aiming to design superalloys with enhanced creep resistance.

1. Introduction

Owing to their colloquial high-temperature mechanical properties, Ni-based superalloys are used in the hot section of gas turbines for energy generation and propulsion. The characteristic high-temperature mechanical properties of such alloys are largely attributed to a potent degree of precipitation strengthening of the γ matrix (disordered fcc) by semi-coherent γ' precipitates ($L1_2$ structure) [1,2]. However, an in depth understanding of the continuous microstructural evolution at high temperature at all the deformation stages, to predict more accurately the materials failure, is still lacking. Reaching a better insight on this topic is essential for the continued optimization of alloy compositions and microstructures, to enhance temperature capability and performance.

The next generation of high-strength, polycrystalline, Ni-based superalloys are pushed towards higher temperature capabilities. Hence, the deformation mechanisms transitions from dislocation glide in the γ channels and either looping around or shearing through the γ' precipitates as paired perfect matrix dislocations separated by an anti-phase boundary (APB) [1,3], to a complex interplay of superlattice intrinsic and extrinsic stacking faults (SISF and SESF), APB, microtwins and stacking faults ribbons [4–11]. These various mechanisms are facilitated, or the resulting deformation features stabilized, by the interaction of the various crystal defects, such as dislocations and stacking faults, with the solutes present in superalloys [12,13].

Some interactions between the crystal defects and the solutes have already been reported. For instance, Co and Cr have been found to segregate preferentially onto partial dislocations along with Re and Mo [14–18]. Segregation at other crystal defects such as APBs, SISFs and SESFs in the γ' precipitates has also been observed [19–23]. In some cases, depending on the level of alloying, these interactions can induce local ordering and result in a phase transformation, such as the formation of ordered, hexagonally-closed-packed (hcp) structured χ and η phases along SISF and SESF, respectively, in some experimental high-refractory content Ni-based superalloys [17,21,24,25]. Quantitative analyses of the local compositional variation at these sub-nanometer features is extremely challenging as the variations are on the order of a few at.% at best and require sensitive analytical techniques. Super Energy dispersive X-ray spectrometry (EDX) – transmission electron microscopy (TEM) has been extensively used providing structural and compositional information from integrated signal through the thickness of the TEM specimen [13,19–24]. More recently, correlative atom probe tomography (APT) and TEM analyses in CoNi-based superalloys have proven to be very efficient to characterize the partitioning of solutes at crystal defects, with near-atomic level quantification of the local compositions [16,17,26].

Although solute partitioning onto crystal defects has become an important factor controlling the mechanical performance of superalloys, quantitative information and fundamental knowledge at the atomic level of such partitioning behaviors remains limited. In some cases, it can be detrimental by promoting either the dissolution of γ' precipitates [14], formation of inverted structures [27], recrystallization [28], or the formation of undesirable topologically close-packed (TCP) phases [29,30]. In other cases, it can be beneficial when the creep strain rate is reduced by a phase transformation mechanism along stacking faults in γ' [21]. Thus, it is of utmost importance to achieve a better understanding of the diffusion-controlled dynamic phenomena leading to the partitioning and precipitation of solutes at crystal defects.

Recent efforts to develop advanced polycrystalline Ni-base superalloys have focused on exploiting increased levels of Ta and/or Nb to improve strength and temperature capability for next-generation disk alloys. In the present study, we investigate the partitioning at crystal defects after creep in an experimental polycrystalline Ni-base superalloys with two different levels of Nb, namely RRHT3 (low-Nb) and RRHT5 (high-Nb), whose compositions are detailed below [31,32]. Their creep performance at the intermediate temperature of 750°C was evaluated. The active deformation mechanisms were imaged by the electron channeling contrast imaging (ECCI) method in a scanning electron microscope (SEM). Finally, the distinct mechanical properties of the two alloys, in terms of creep strain rate, are rationalized by an in-depth APT study of the interaction between the alloying elements and the crystal defects created by deformation. Based on the new knowledge gained from our study, an approach for alloy design is suggested, to better target the phase transformation on stacking faults that leads to improved creep properties.

## 2. Materials and methods

Two alloys with a composition given in Table 1 and that are referred to as RRHT3 (low-Nb) and RRHT5 (high-Nb) were studied. The compositions of these alloys are nominally identical apart from the Al and Nb levels, both being γ' forming elements. The alloys were hot isostatically pressed (HIPed), isothermally forged and solution treated at 1170°C for 1 hour, and cooled down to room temperature at a controlled rate of 1°C/s. An ageing treatment was then performed at 850°C for 4 hours, followed by furnace cooling. Standard tensile creep samples with a gauge length of 50mm and a gauge diameter of 5mm were crept at 750°C and 600 MPa in air. The creep strain in the samples was measured using standard ASTM E139 creep test methods. Strain was measured using linear variable differential transformers (LVDTs) attached to the extension arms that were fixed onto the collars of the creep specimens. This method measures the overall (uniform and non-uniform) creep strain during the test. The specimens tested exhibited relatively little ductility prior to failure and the accumulated strain was largely uniform in nature.

Pre- and post-creep metallographic samples were cut along the loading direction, grinded with SiC papers and polished with 0.04µm colloidal silica. The fully heat-treated samples were etched in a 33%HCL −33%NO3 −33%CH3COOH −1%HF solution, and high-resolution microstructural observations were carried out on a JEOL JSM 6701-F field emission scanning electron microscope (FESEM), with an accelerating voltage of 10 kV and current of 10 µA. ECCI analysis of the crept microstructures was performed on a Zeiss Merlin SEM equipped with a backscattered electron (BSE) detector, at an accelerating voltage of 20 kV and a probe current of 2 nA.

SEM-BSE imaging was used to locate grains with a high density of defects in the samples fractured in creep. These grains were targeted for APT specimen preparation by using a dual beam SEM/focused ion beam (FIB) instrument (FEI Helios Nanolab 600i) via a site-specific lift-out protocol outlined in Ref. [33]. APT measurements were carried out on a Cameca LEAP$^{TM}$ 5000 XR operated in laser pulsing mode at a pulse repetition rate of 125 kHz and a pulse energy of 55 pJ. The base temperature was set to 60K and the detection rate was maintained at 15 ions every 1000 pulses. Data analyses were performed using the IVAS 3.8.4 software package. APT measurements of the fully heat treated microstructures were performed on a LEAP 4000$^{TM}$ XSi operated in laser pulsing mode at a pulse repetition rate of 500 kHz, a pulse energy of 25 pJ, a pulse repetition rate of 500 kHZ, a base temperature of 34 K and an evaporation rates of up to 15 ions every 1000 pulses.

*Table 1: Nominal composition of the RRHT3 and RRHT5 alloys (at.%)*

|  | Ni | Al | Co | Cr | Mo | Nb | Ta | W |
|---|---|---|---|---|---|---|---|---|
| RRHT3 | Bal. | 9.8 | 18.3 | 5 – 16 | 0 - 3 | 4.6 | 1.0 | 0 - 2 |
| RRHT5 | Bal. | 8.0 | 18.1 | 5 - 16 | 0 - 3 | 5.5 | 1.0 | 0 - 2 |

3. **Results**

   3.1. **Heat-treated microstructure and creep performance**

The initial microstructure of the two studied alloys after heat treatment and prior to creep deformation is shown in Figure 1. The two alloys exhibit the typical γ/γ' microstructure observed in most superalloys. Besides, γ-like precipitates in the γ', originating from the supersaturation of Cr and Co in γ', are also evidenced in the insets of Figure 1a and b.

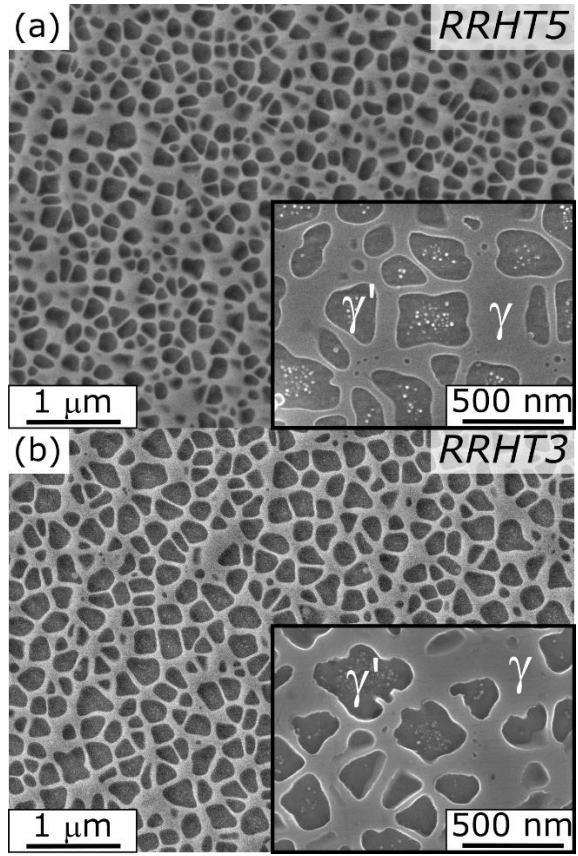

*Figure 1: Micrographs of the fully heat treated (a) RRHT5 and (b) RRHT3 alloy. The low magnification pictures are imaged with secondary electrons and the insets at higher magnification are imaged with in-lens technique.*

Using the composition of the γ/γ' microstructure as measured by APT for both alloys after full heat treatment, the volume fraction of γ' precipitates was calculated. In particular, we plotted the difference in composition between the nominal concentration for each element and the solute content in the γ matrix ($C_n$–$C_\gamma$) as a function of the difference in composition between γ' and γ ($C_{\gamma'}$- $C_\gamma$) following [34]. The volume fraction of γ' corresponds to the slope of the best-fit line showing in Figure 2, which is 53.9% for the RRHT3 alloy and 55.7% for the RRHT5 alloy. This method has been shown to provide reliable results, in good agreement with thermodynamic predictions and SEM analyses [35].

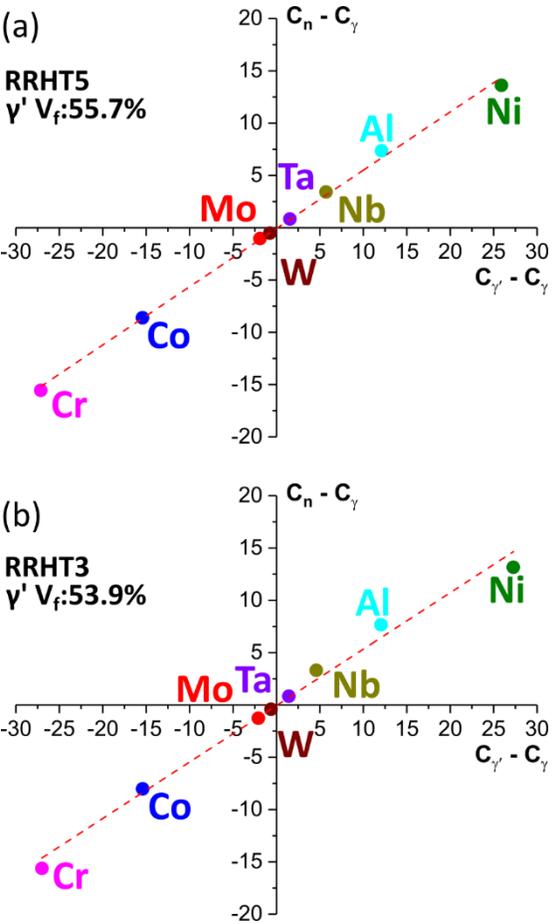

*Figure 2: Lever rule graph used to calculate the γ' volume fraction after full heat treatment for RRHT5 (a) and RRHT3 (b) alloys.*

### 3.2. Creep performance and deformation at 750°C

The creep curves at 750°C and 600 MPa of the two alloys are displayed in Figure 3. RRHT5 exhibits a significantly reduced steady-state creep strain rate of $2.82 \times 10^{-6}$/h when compared to RRHT3, $8.20 \times 10^{-6}$/h. Additionally, the time to 0.5% creep strain for RRHT3 and RRHT5 was 159h and 277h, respectively. The RRHT3 creep sample ruptured following ~480h at these test conditions while the RRHT5 creep sample fractured after ~380h.

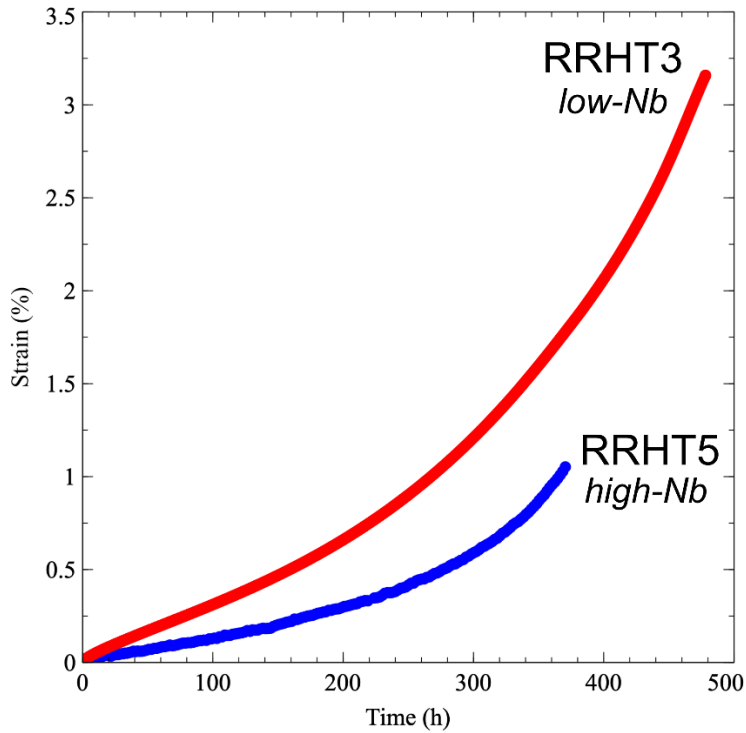

*Figure 3: Creep strain vs. time curves for RRHT3 (in red) and RRHT5 (in blue) of samples tested at 750°C under a 600 MPa load.*

ECCI analyses of the deformed microstructures are given in Figure 4, and close ups are provided in Figure 4b and 4d for the Figure 4a and 4c, respectively. The provided micrographs are representative of the deformed microstructure observed in the two considered alloys. Stacking faults are visible in RRHT5 (Figure 4a) along three directions in the grain, highlighted in Figure 4b by green, blue and red lines. These features can be identified as stacking faults by their extinction fringe contrast and as previously reported for other superalloys [17,36]. The RRHT3 alloy also displays bright elongated planar features in the γ' precipitates. All share the same direction. Some are interrupted at the interface between the precipitate and the matrix (see red arrow in Figure 4d) and are also identified as stacking faults, while others extend across the matrix and the precipitates, as highlighted by the red circle in Figure 4d. It is suggested that the latter are microtwins based on similar observations in other alloys [12,37]. One can note that microtwins are not observed in RRHT5.

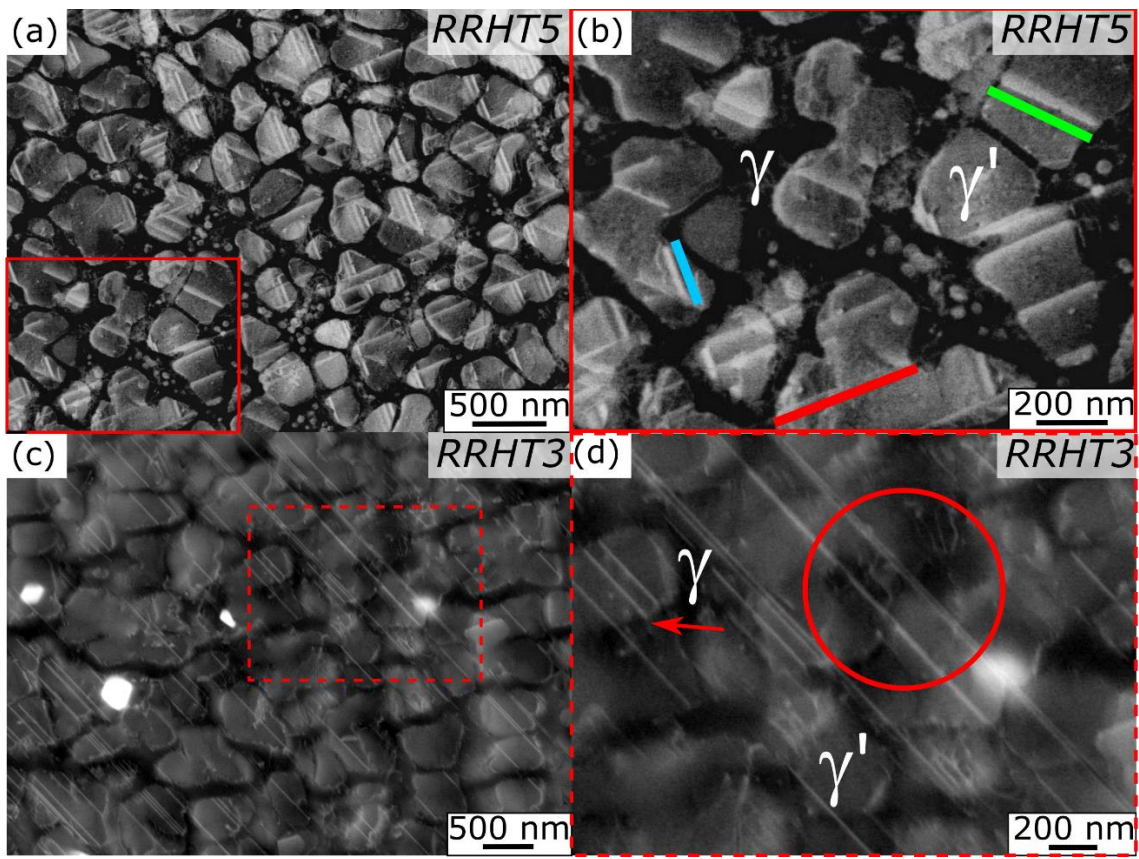

*Figure 4: (a) ECCI image of the microstructure of RRHT5, **fractured in creep** and (b) Magnified image of the red plain rectangle in (a), highlighting three different orientations for the stacking faults in blue, red and green lines. (c) ECCI image of the microstructure of RRHT3, fractured in creep. (d) Magnified image corresponding to the red dashed rectangle in (c), the red arrows highlight stacking faults that are interrupted in at the γ/γ' interface, and the red circle shows microtwins that extend across the γ matrix.*

### 3.3. **Partitioning of solutes at stacking faults in RRHT5**

Given the slower creep strain rate of the RRHT5 alloy, our efforts have been focused first on understanding the interactions of solutes with stacking faults in this alloy.

Figure 5 shows an APT analysis of the RRHT5 alloy (high-Nb), fractured in creep. The tomographic reconstruction in Figure 5a includes the γ matrix and part of a γ' precipitate. A 3 at.% Cr iso-composition surface is superimposed onto the point cloud. It highlights the interface between γ and γ', as well as a tubular feature connected to the interface (see red arrow #1). Similar features appearing in analyses performed on several metallic materials, including other Ni-based or Co-based superalloys allow us to identify it as the segregated Cottrell-type atmosphere at a dislocation [14,38–40]. It also highlights a round secondary γ particle inside the γ' precipitate on the left-hand side (red arrow #2), a dislocation (tubular feature visualized along the tube's main axis, see red arrow #3)) and a plane behind it. 2D composition maps are plotted in Figure 5b and 6c for Cr and Co, respectively, and reveal the details of the segregation behavior in the region of the dislocation followed by a plane, that was visualized in Figure 5a and delimited by a dashed black rectangle.

The 2D composition map of Figure 5b confirms that the dislocation is enriched in Cr, and followed by a Cr-rich plane. The Co 2D-composition map of Figure 5c also shows that the dislocation and the plane attached are enriched in Co, and evidences the presence of a second plane parallel to the plane associated with the dislocation, which is also enriched in Co. This second plane is not related to the dislocation, and extends across the entire reconstructed dataset.

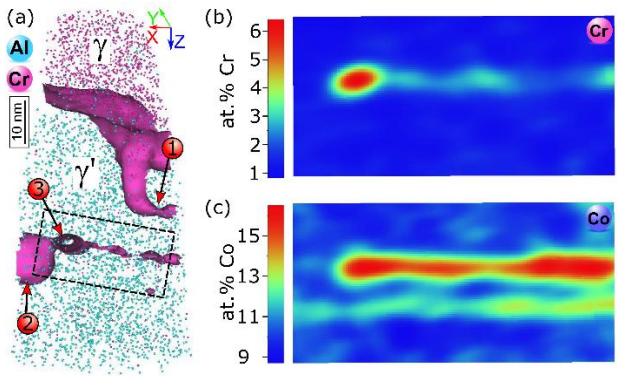

*Figure 5: APT analysis of the crept RRHT5 alloy. (a) atomic reconstruction displaying the γ matrix (at the top) and a γ' precipitate (bottom). The interface is represented by a Cr iso-composition surface at 3 at.%. (b) and (c) 2D composition profiles for Co and Cr, respectively, corresponding to the region represented by the black dashed box in (a).*

The composition of the dislocation is determined by the 1D composition profiles (along arrow #1 in Figure 6a) given in Figure 6b and 6c. As highlighted by a purple background at the location of the dislocation, a strong segregation of Co and Cr is observed, alongside with a small Mo segregation, while Nb is depleted. Similar segregation has been observed in other superalloys, however, to the best of our knowledge, the depletion of Nb at dislocations has not been reported before.

These composition profiles also provide information for the second observed plane (at around 9nm, indicated by a grey background on the graphs), which is not associated with the dislocation. In particular, the second plane is enriched in Co and Nb, with no apparent segregation or depletion of Cr and Mo. 1D composition profiles were calculated at a different location for both planes as indicated by the green arrow #2 in Figure 6a. Cr and Mo, plotted in Figure 6d, show an enrichment in the first plane up to 2.9 and 1.4 at.%, respectively, but no clear enrichment in the second plane. Co and Nb, plotted in Figure 6e, show a clear enrichment up to 18.3 and 9.8 at.%, respectively, in the first plane. As already noticed in the plots of Figure 6b and 6c, and contrary to Cr and Mo, Co and Nb are also segregated in the second plane, but to a lesser extent with values up to 13.4 at.% for Co and 9.0 at.% for Nb. The difference in partitioning between those two planes suggests the presence of two planar faults with different character, and their relative identification is discussed in more details in the discussion section.

Finally, a 1D composition profile was extracted from a cylindrical region taken along the first plane, and is plotted for Cr, Co and Nb in Figure 6f. An enrichment in Co and Nb and to a lesser extent in Cr, in the SISF (blue region) compared to the composition of the γ' ahead of the plane of the fault is shown. Together with the composition profiles plotted across the planes, this suggests that diffusion of solutes to the planar fault occurs along the plane and not perpendicular to the plane. This is also supported from similar observations in a Co-based superalloy [17].

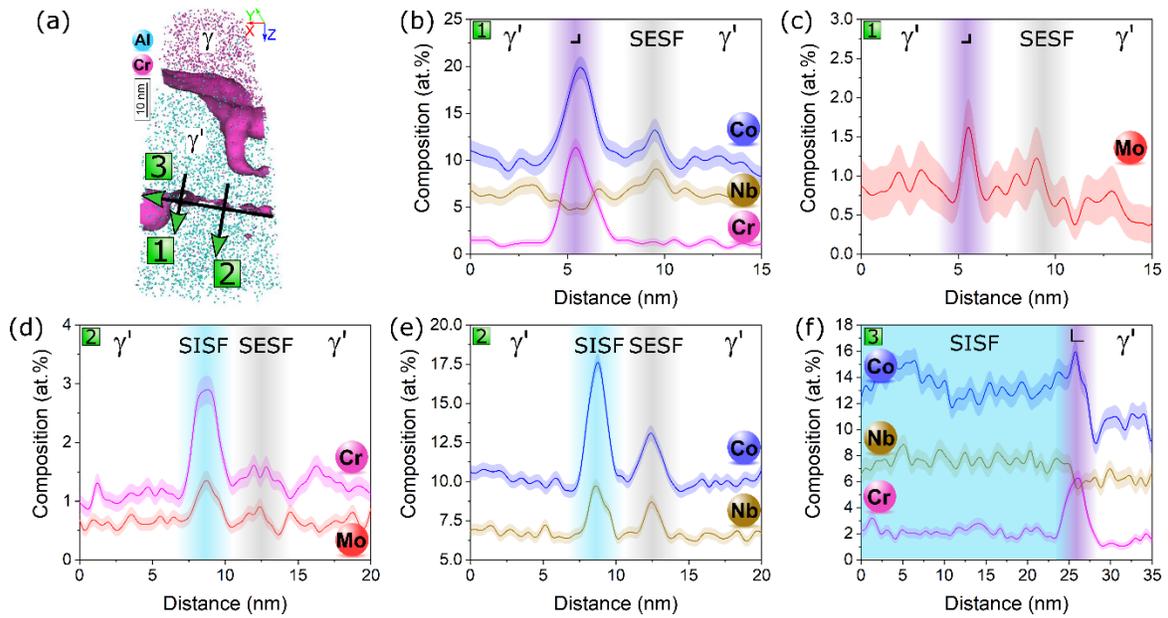

*Figure 6: APT analysis of the crept RRHT5 alloy. (a) atomic reconstruction from Figure 5a. (b) and (c) 1D composition profiles calculated along the green arrow #1 in (a) for Cr, Co and Nb and for Mo, respectively. (d) and (e) 1D composition profiles calculated along the green arrow #2 in (a) for Cr and Mo and for Co and Nb, respectively. (f) 1D composition profiles calculated along the green arrow #3 in (a) for Cr, Nb and Co. Error bars are shown as lines filled with colour and correspond to the 2σ counting error.*

Similar Nb depletion at the dislocation is more clearly evidenced in Figure 7, where a 1D composition profile is plotted across a Co-, Cr-rich dislocation (shown by the yellow arrow in Figure 7a).

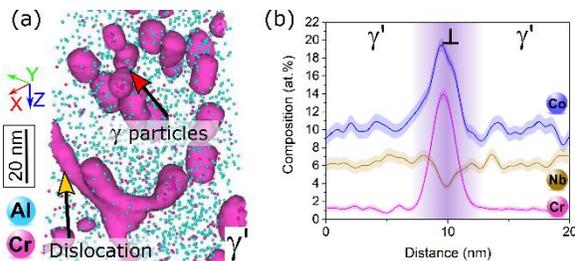

*Figure 7: APT analysis of the crept RRHT5 alloy. (a) atomic reconstruction of a γ' precipitate containing many tertiary γ. Cr iso-composition surface at 3.2 at% is plotted. A dislocation is visualized (yellow arrow). (b) 1D composition profile extracted across the dislocation in (a)*

*for Co, Nb and Cr. Error bars are shown as lines filled with colour and correspond to the 2σ counting error.*

Such results are obtained repeatedly in various dataset of the RRHT5 alloy fractured in creep, as shown in Figure 8. The Cr, Nb and Co 2D-compositions profiles of Figure 8b, 8c and 8d, respectively, are extracted from the box shown in the atomic reconstruction of Figure 8a. They highlight the presence of a Cr, Co enriched dislocation (red arrow) depleted in Nb. Two Co, Nb rich parallel planes are also observed. A 1D composition profile across the top Nb, Co-rich plane and the dislocation confirms (Figure 8e and 8f), similar to the results of Figure 6, that the dislocation is enriched in Co and Cr, but also in Mo, and depleted in Nb and in Ta. The SESF plane is enriched in Co and Nb, as suggested by the 2D composition profiles, and also slightly in Mo.

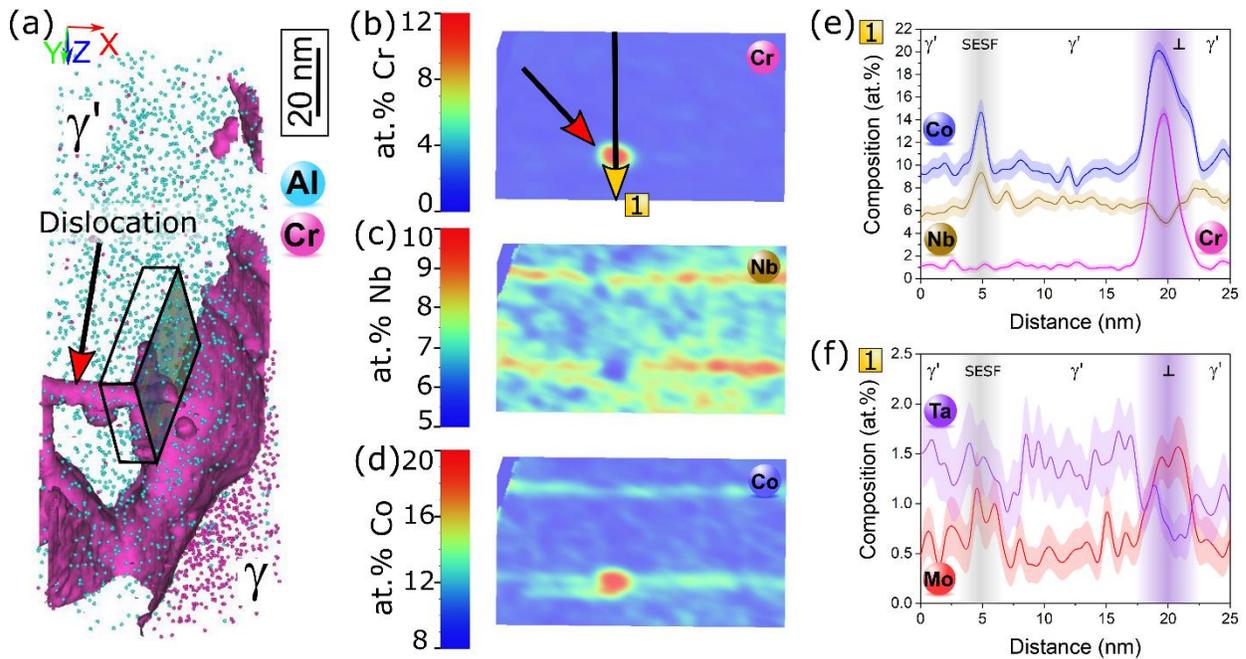

*Figure 8: APT analysis of the crept RRHT5 alloy. (a) atomic reconstruction displaying the γ matrix and a γ' precipitate. The interface is represented by a Cr iso-composition surface at 2.8 at%. A dislocation is visualized (red arrow). (b), (c) and (d) 2D iso-composition surface for Cr, Nb and Co, respectively, extracted from the box in (a). The red arrow in (b) highlights the location of the Cr-rich dislocation. (e) and (f) 1D composition profile extracted along the*

*yellow arrow #1 in (a) for Co, Nb and Cr and for Ta and Mo, respectively. Error bars are shown as lines filled with colour and correspond to the 2σ counting error.*

Figure 9 shows part of another APT analysis of the RRHT5 alloy. Once again, γ and γ' are visualized, along with two dislocations that have intersected the γ/γ' interface (Figure 9a). As indicated by the 2D composition profile in Figure 9b, the two dislocations are enriched in Co, and the plane between them is also enriched in Co, although to a lesser extent, compared to the rest of γ'. The 1D composition profile (Figure 9c) extracted from a cuboidal region of interest along the yellow arrow (in Figure 9a) shows in particular that for one dislocation, Cr reaches up to 9.0 at.% and Co up to 20.4 at.%. Although the same solutes partition at the second dislocation, their amount is lower, with Cr up to 8.2 at.% and Co 14.9 at.%.

Therefore, it is proposed as an hypothesis that the two dislocations are a pair of partial dislocations. In particular, this is suggested based on the presence of a faulted plane delimited by these two dislocations. Following the higher amount of segregation at one of the two dislocations, the latter is identified as the leading partial dislocation (LPD), while the second one is identified as the trailing one. However, this identification is, at this point, a suggestion that merely helps naming the dislocations in the following, and does not have an impact on the proposed mechanism. Similar differences between partials have been previously observed and calculated in nickel- and cobalt-based superalloys [14,16,28,29,41–43].

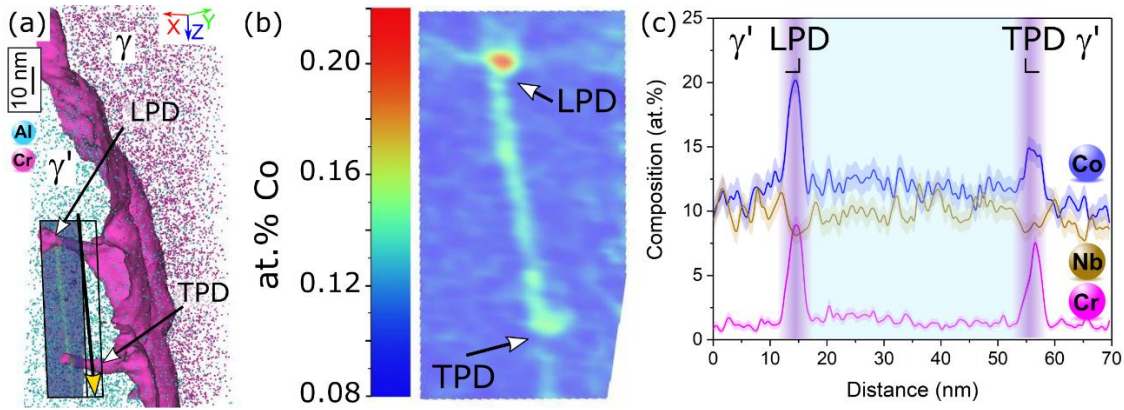

*Figure 9: APT analysis of the crept RRHT5 alloy. (a) atomic reconstruction displaying the γ matrix and a γ' precipitate. The interface is represented by a Cr iso-composition surface at 3 at%. A leading partial dislocation (LPD) and a trailing partial dislocation (TPD) are visualized. (b) 2D Co iso-composition surface extracted from the box in (a). (c) 1D composition profile extracted along the yellow arrow in (a), showing the two partials and the stacking fault in between. Error bars are shown as lines filled with colour and correspond to the 2σ counting error.*

The APT volume of Figure 9a was rotated about its z-axis, leading to Figure 10a. This time, the partials are perpendicular to the observation plane. The 2D composition profile for Co was plotted perpendicular to the LPD, and is displayed in Figure 10a. It shows that the LPD is situated at the crossover of four Co-enriched planes, numbered from 1 to 4. Plane #1 extends between the LPD and TPD and the partitioning of solutes there (Figure 10b and 10c), compared to the partitioning at the other planes (Figure 10d, 10e and 10f), is also indicative that it corresponds to a SISF. Indeed, the 1D composition profiles from a cylindrical region of interest across the plane #1 clearly evidence partitioning of Cr, Co, Nb and Mo (Figure 10b and 10c), similarly to the upper plane behind the dislocation in Figure 5. An enrichment in Co and Nb (15.1 at.% and 11.3 at.%, respectively) is measured in this plane, and segregation of Cr and Mo reaches levels up to 2.2 and 1.2 at.%, respectively.

By contrast, in the other planes, only segregation of Co and Nb is observed, and no enrichment in Cr or Mo was evidenced there, suggesting that these planes are SESF. For instance, the composition profile across plane #2 (Figure 10d) highlights an enrichment in Co (up to 14.5 at.%) and in Nb (up to 12.2 at.%). Co and Nb are also segregated in the plane #3 (Figure 10e), but in a lower amount than for the other planes #1, 2 and 4. Finally, a composition profile plotted across the plane #4 (Figure 10f) highlights two successive increases in Co and Nb, at a distance of 5 nm from each other. Only the elements segregated are included in these plots, meaning that when no increase of Cr or Mo was measured, these elements were not plotted.

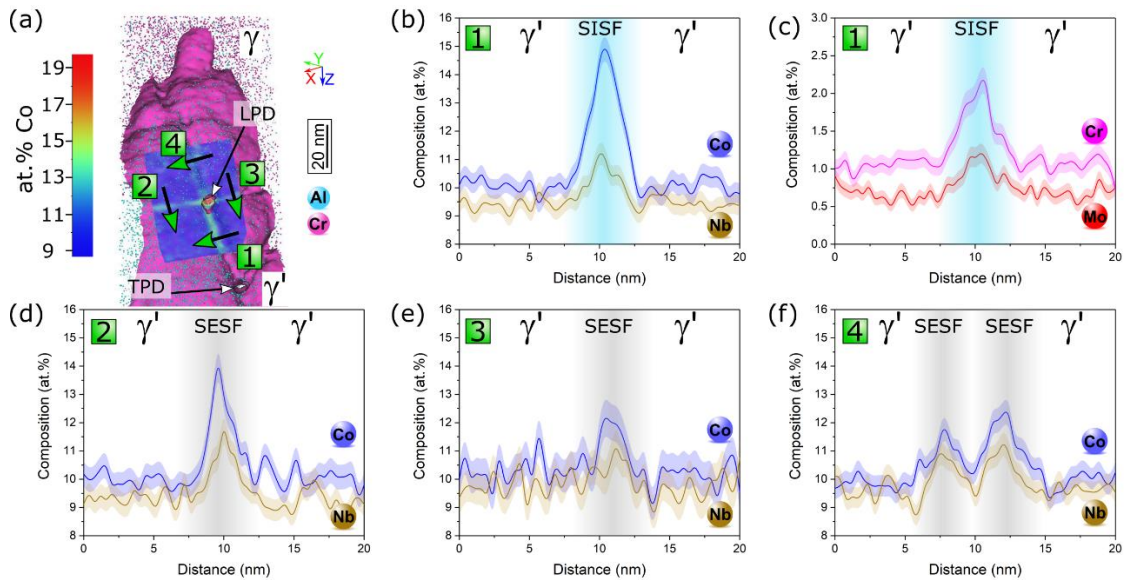

*Figure 10: APT analysis of the crept RRHT5 alloy. (a) atomic reconstruction of the same volume as in Figure 9a but rotated along z, along with a 2D compositional map for Co. The γ/γ' interface is represented by a Cr iso-composition surface at 3 at%. (a) to (f): 1D composition profiles (a) for Co and Nb, across the plane #1 in Figure 10a, (b) for Cr and Mo across the plane #1. (d) (e) and (f) for Co and Nb across the planes #2, #3 and #4 in Figure 10a, respectively. Error bars are shown as lines filled with colour and correspond to the 2σ counting error.*

Cr and Mo enrichment at SISF is last shown in Figure 11, confirming repeatability of the observations. In the presented dataset, two dislocations are observed, highlighted in the reconstruction by the Cr isocomposition surface, and two SISF are visualized in neighboring parallel planes. 1D composition profiles plotted along the green arrow (in Figure 11b) are given in Figure 11e and 11f, and confirm the segregation of Nb, Co, Cr and Mo at the SISF.

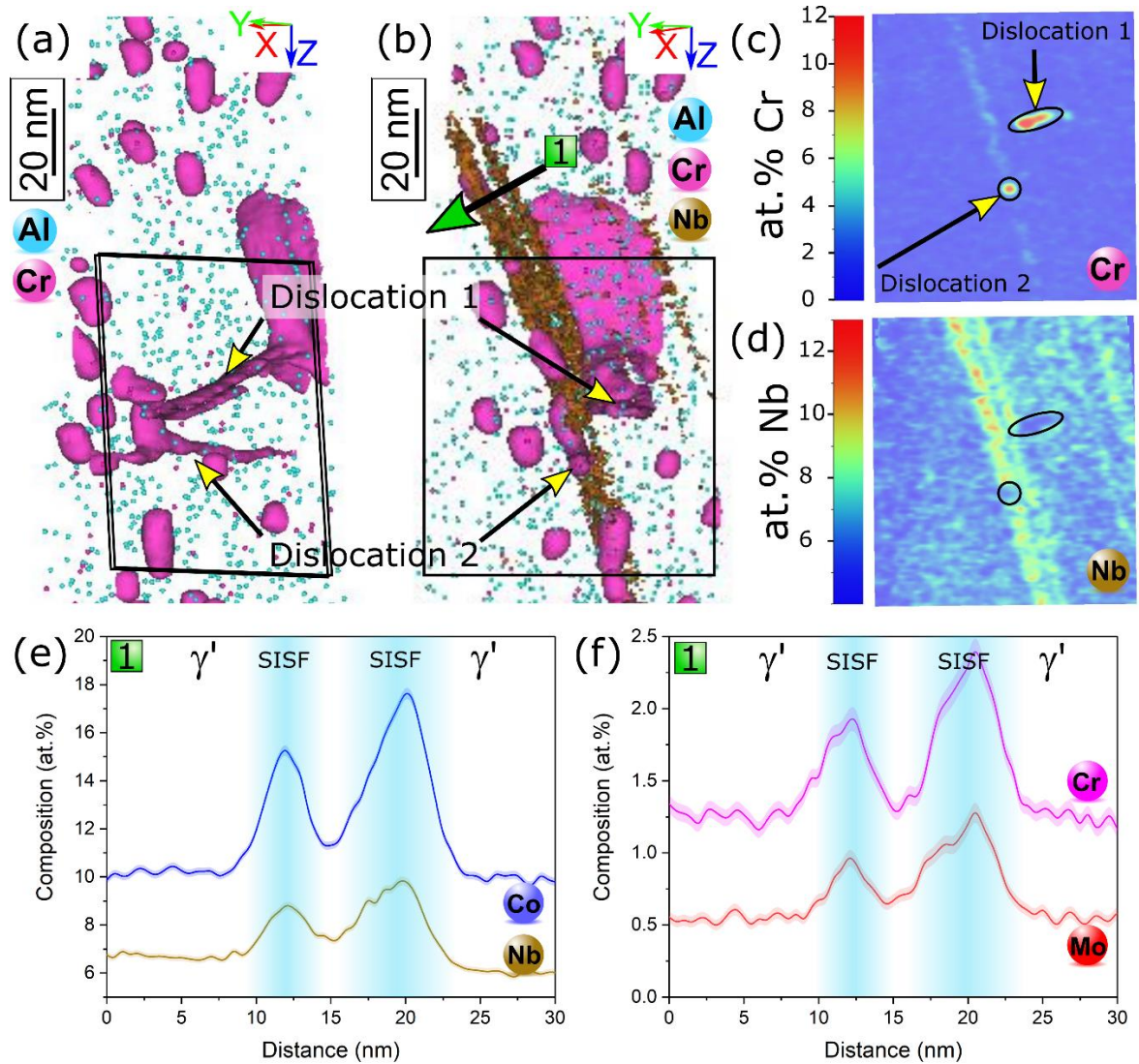

*Figure 11: APT analysis of the crept RRHT5 alloy. (a) and (b) atomic reconstruction of a γ' precipitate. Cr and Nb iso-composition surface at 4.5 at% and 9 at.%, respectively, are plotted (Nb isocomposition surface plotted in brown only in (b)). Two dislocations are visualized (yellow arrows). (c) and (d) 2D iso-composition surface for Cr and Nb, respectively, extracted from the box in (a). (e) and (f) 1D composition profile extracted along*

*the green arrow #1 in (b) for Co and Nb and for Cr and Mo, respectively. Error bars are shown as lines filled with colour and correspond to the 2σ counting error.*

### 3.4. Atomic investigation of the deformation mechanisms in RRHT3

Although in the case of RRHT3 alloy (low-Nb), the presence of microtwins is well known to accelerate deformation [12,44], details of the interplay between solute and deformation mechanisms were also investigated by APT performed on the fractured sample.

Figure 12a shows a selected region of an APT dataset from the crept-to-fracture RRHT3 alloy. An interface between a γ' precipitate and the γ matrix is visualized by a 10.5 at.% Cr isocomposition surface. A dislocation is connected to the interface and extends into γ'. Similarly to RRHT5, composition profiles in Figure 12b to 12d highlight an enrichment in Co, Cr and Mo at the dislocations. By contrast, Nb and Ta are depleted (Figure 12c). The segregation at the dislocation is slightly higher in RRHT3 than in RRHT5: Co is segregated up to 22.3 at.%, Cr 18.0 at.% and Mo 1.7 at.% for RRHT3. The fault in-between is enriched in Co, Cr and Mo.

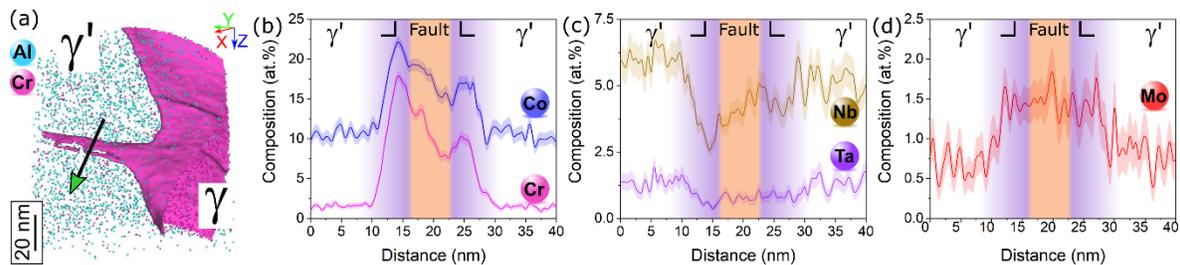

*Figure 12: APT analysis of the crept RRHT3 alloy. (a) APT reconstruction displaying a γ' precipitate on the left and the γ matrix on the right. The interface is represented by a Cr iso-composition surface at 10.5 at%. The composition profile along the green arrow is plotted in (b) for Co and Cr, in (c) for Nb and Ta and in (d) for Mo. Error bars are shown as lines filled with colour and correspond to the 2σ counting error.*

Finally, as shown in the Figure 13, RRHT3 also contains faults with segregations. The 2D composition map provided for Co in the reconstruction of Figure 13b highlights the presence of a Co-rich plane in the γ' precipitate. Composition profiles for Co (Figure 13c), Nb (Figure 13d) and Cr and Mo (Figure 13e) were extracted from a cylindrical region of interest across this plane (along the green arrow in Figure 13b). They reveal that the plane is, as in RRHT5, enriched in Co, Nb, Cr and Mo. Therefore, as proposed in the discussion, it is also identified as an SISF.

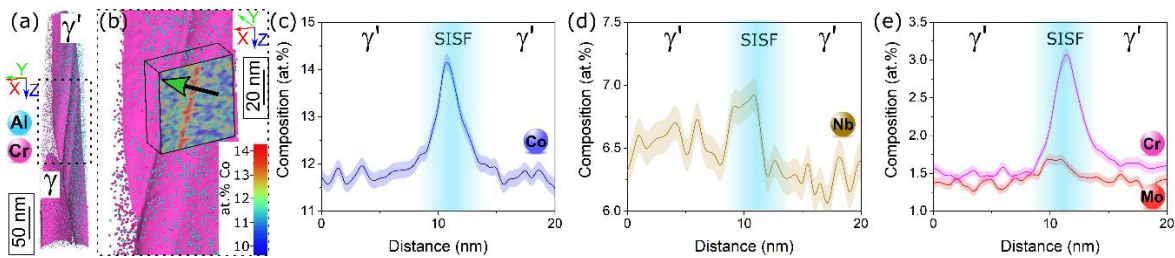

*Figure 13: APT analysis of the crept RRHT3 alloy. (a) full APT reconstruction displaying a γ' precipitate in the front and the γ matrix in the back. The interface is represented by a Cr iso-composition surface at 10.0 at%. (b) Higher magnification of the region in a dashed bow in (a), along with a Co 2D composition profile. The composition profile along the green arrow in (b) is plotted in (c) for Co, in (d) for Nb and in (e) for Cr and Mo. Error bars are shown as lines filled with colour and correspond to the 2σ counting error.*

In overall, APT specimen originating from two distinct lift-outs for each alloy were analyzed, for a total number of 9 samples studied. This ensures that the observations reported in this paper are reliable. A total of 13 dislocations, 17 SESF and 4 SISF were obtained and characterized, allowing to propose the results presented here.

## 4. Discussion

The results presented above highlight better creep properties for a Nb-rich alloy, RRHT5, compared to a similar alloy with a 0.9 at.% lower nominal Nb content, RRHT3. The reasons of the improved creep performance of RRHT5 will first be discussed, and the outcomes will be integrated in a second part into a proposed design tool, aiming at predicting precipitation onto stacking faults, and thus improved creep properties for Ni-based superalloys.

### 4.1. Atomic-scale phase transformation for enhanced creep performance

Several parameters can influence the creep properties, such as the initial microstructure or the composition. Regarding the two studied alloys, based on the SE images of Figure 1, showing similar initial microstructures, and the related quantification of the amount of γ' precipitates (Figure 2), it is reasonable to believe that the small difference in the γ' volume fraction (being 3% less for RRHT5 than for RRHT3) cannot explain the difference in creep behavior.

Therefore, the compositional difference between RRHT3 and RRHT5 was considered to rationalize the distinct creep behaviors. Table 1 indicates that there is 0.9 at.% more Nb and 1.8 at.% less Al in the nominal content of RRHT5 compared to that of RRHT3. As a consequence, the partitioning between γ and γ' will be affected, resulting in γ' precipitates with different composition. Table 2 shows the composition of γ' as measured by APT for both alloys (prior to creep deformation, traces of other elements are not provided). As indicated in Table 2, although most of elements partition with the same amount in RRHT3 and RRHT5, there is a noticeable difference in the amount of Nb, which is 1.0 at.% higher in RRHT5 alloy compared to RRHT3 alloy.

*Table 2: Summary of composition in at.%. of the γ' phase as measured by APT in both alloys after heat treatment.*

|       | Ni | Al | Co | Nb | Ta | Cr | W | Mo |
|-------|-----|-----|-----|-----|-----|-----|-----|-----|
| RRHT3 | 63.75 ± 0.03 | 14.77 ± 0.02 | 10.78 ± 0.02 | 5.97 ± 0.01 | 1.58 ± 0.01 | 1.94 ± 0.01 | 0.430 ± 0.003 | 0.508 ± 0.004 |
| RRHT5 | 63.67 ± 0.02 | 14.17 ± 0.01 | 10.89 ± 0.01 | 6.96 ± 0.01 | 1.525 ± 0.005 | 1.886 ± 0.005 | 0.387 ± 0.002 | 0.428 ± 0.003 |

Identification of the deformation mechanisms was done by ECCI. It shows that both alloys deform by dislocation glide and that partial dislocations separated by stacking faults are observed within the precipitates. Additionally, microtwinning is evidenced for RRHT3.

Considering the similar initial microstructures, and the presence of stacking faults in both alloys, interpretation of the different creep behavior was thus sought on the partitioning side of elements at deformation defects, resulting from the γ' compositional change between the two alloys.

APT analyses of crept samples reveal for both alloys segregation of Co, Cr and Mo at dislocations (see Figure 5 and Figure 12). Stacking faults enriched in Co, Nb, Cr and Mo are also observed by APT in both alloys. Yet, a second type of planar faults, enriched in Co and Nb only, is observed in RRHT5. As shown in Figure 9, the planes of the different faults may intersect in the RRHT5 structure: in the considered case, the LPD interacts with four planes other than its own stacking fault plane. Similar observation of intersecting planes was reported by Barba et al. [22]. They evidenced, with help of STEM coupled with EDX, the intersection of several superlattice extrinsic stacking faults (SESF) and of a superlattice intrinsic stacking fault (SISF). Their analysis shows two SESF lying approximately 10nm away from each other, in a plane that has the same orientation as the SISF, similar to the present APT study, the planes #4 (the two SESF) and the plane #1 (the SISF) in Figure 9a. EDX measurements on the single crystal MD2 alloy, crept at 800°C under a 650 MPa load, revealed an enrichment in γ-stabilizers, such as Co, Cr and Mo at both SESF and SISF [22].

However, other EDX studies on different crept superalloys reported that the segregation behavior may differ from one type of fault to another [19,21,24]. As an example, the LSHR alloy with high Nb and W content and a rather low Ta and Mo content, was shown to display Co, W, Nb, Cr and Mo segregation onto the SISFs, that is believed to evolve to a χ $Co_3$(W,Cr,Mo)-type phase [19,21]. The SESFs in LSHR are enriched in Co, Nb, W and Mo, and, in a lesser extent, in Ta and Ti, and the η-$Ni_3$(Nb,Ta) phase is believed to form [21,24]. By comparing the RRHT5 alloy with LSHR, the partitioning of Cr at a fault would be the signature of a SISF, while its absence (with only the segregation of Co and Nb) would indicate that the fault is a SESF. This would also be in good agreement with the results of Barba *et al*. [22] that show that Cr segregation at SESF is lower than that at SISF. These observations also line up with the formation mechanism of SISFs, that originate from an antiphase boundary (APB) and a complex stacking fault (CSF) [45]: calculations showed that CSFs are strongly stabilized by substitution on Ni sites of Nb and Ta (approximately -80 mJ/m$^2$), by Cr (approximately -60 mJ/m$^2$) and by Co (about -10 mJ/m$^2$) and APB are stabilized by substitution on Ni sites by Co, Cr, Nb and Ta (approximately -10 mJ/m$^2$, -75 mJ/m$^2$ for Co and Co, respectively, and -100 mJ/m$^2$ for Nb and Ta) [20]. Therefore, Co, Cr and Nb should all be found in SISF.

Based on these results from the literature, it is proposed that SISFs, with segregation of Co, Cr, Nb and Mo are found in RRHT3 and RRHT5, and SESFs with segregation of Co and Nb are only found in RRHT5. Furthermore, the compositional trends observed in the present study, compared to the published studies discussed above, suggest that precipitation onto the faults could occur, leading to the formation of the ordered η-phase on the SESF and of the χ-phase on the SISF.

The discrepancy between the two alloys would therefore rely on their ability to form η-phase, with D0$_{24}$ structure, or its precursor, on the SESF, as in RRHT5. Indeed, stacking faults ribbons, commonly observed in superalloys crept in these temperature ranges are made of a SISF + APB + SESF sequence [5]. If, after diffusion of the segregated elements, a local phase transformation occurs on the SESF following the glide of the leading partial that transforms it to a local η phase, further shearing of adjacent partials, i.e. microtwin formation, would be energetically unfavorable and effectively suppressed, which improves the creep properties, as was already suggested in earlier work [21]. Without such segregation favoring local η phase formation, RRHT3 is more prone to forming microtwins by additional partials shearing adjacent to the SESF planes. Additionally, diffusion of heavy atoms like Nb to the SESF would likely slow down the creep rate further.

At last, this diffusion of solutes to the deformation features could occur by pipe diffusion along the LPD, diffusion along the fault (in plane diffusion) or from the rest of the γ' precipitate, with a diffusion occurring perpendicular to the fault [14,17,25,46]. Here, diffusion occurring perpendicular to the fault is excluded: none of the composition profiles plotted across a fault showed a depletion of solutes above or below the fault plane. However, Figure 5h shows that Nb and Co are both depleted ahead of the LPD, Co is enriched at the LPD and Nb is depleted, and both are enriched in the following SISF. As for Cr, it is enriched at the LPD and, to a lesser extent, at the fault. Therefore, it is suggested that a combination of pipe-diffusion and in-plane diffusion occurs for Cr and Co, and only in-plane diffusion occurs for Nb. Cr and Co would be "sucked" by the LPD from the region ahead, and remain in the SISF, in the wake of the LPD. However, the absence of Cr depletion ahead of the LPD suggests that occurrence of pipe diffusion cannot be dismissed.

As for Nb, its tendency to be depleted at the dislocation core rules out the hypothesis of pipe diffusion, and so Nb in SISF must come from in-plane diffusion. These results show that in-plane diffusion occurs in this alloy, probably assisted by pipe diffusion for Co and Cr. Therefore, mechanisms distinct for the various solutes but working in a collaborative manner to produce the observed segregation.

### 4.2. Design guideline for phase transformation along faults

It is striking that the small compositional difference between the two alloys, being 0.9 at.% for Nb and 1.8 at% for Al, is responsible for such a considerable change in the deformation mechanism, and more specifically the segregation tendency to the faults.

To rationalize the difference in creep behavior between the two alloys and gain some insight on how to design future alloys by exploiting such effects, an attempt was made to create a first design diagram, in Figure 14, based on the relatively few results found in literature. Smith et al. [44] pointed out that the superior creep resistance of ME501, compared to that of ME3, was the result of a local structural and chemical transformation of SESFs to η-phase.

Hence, the design goal would be to promote such local transformations to ordered, hard, geometrically closed packed (GCP) phases of the $Ni_3X$ nature, especially at SESFs. The formation of ordered η-phase on a SESF would highly depend on the availability of Nb and Ta atoms in the γ' phase, and so high concentrations of these γ' forming elements seem to be desirable. However, exceeding the alloys solubility limit would promote precipitation of bulk η and/or δ [47] in the alloy, once again resulting in creep performance deterioration. Therefore a careful balance of γ' forming elements needs to be made such that the alloy lays in a metastability region that has the maximum amount of Nb, Ta and Ti relative to Al, but does not exceed a limit which would make another ordered phase more stable than γ'.

The γ' composition was used for the construction of Figure 14 instead of the bulk composition since the solubility of γ' changes with temperature, which implies that the heat treatment of the alloy can also influence the deformation fault nature to some extent. Further, as was shown in [47], Ta and Nb are more potent GCP stabilizers than Ti, hence a potency weight factor of 0.5 was assigned to the Ti content for Figure 14. Based on the relatively few data available in literature, it would seem that whenever the (Nb+0.5Ti+Ta) composition exceeds ~0.625 times the Al content, a local η-phase phase transformation at the SESF is promoted. It should be noted that, the empirical nature of these fitting numbers requires large datasets for good accuracy, however, the challenge of extracting such details during characterization has resulted to knowledge on only a handful of alloys [21,22,25,48]. As more alloys are explored, the exact limits of how much Nb, Ta and Ti need to be present for local (and avoiding bulk) η-phase formation can be refined and more creep resistant alloys can be designed.

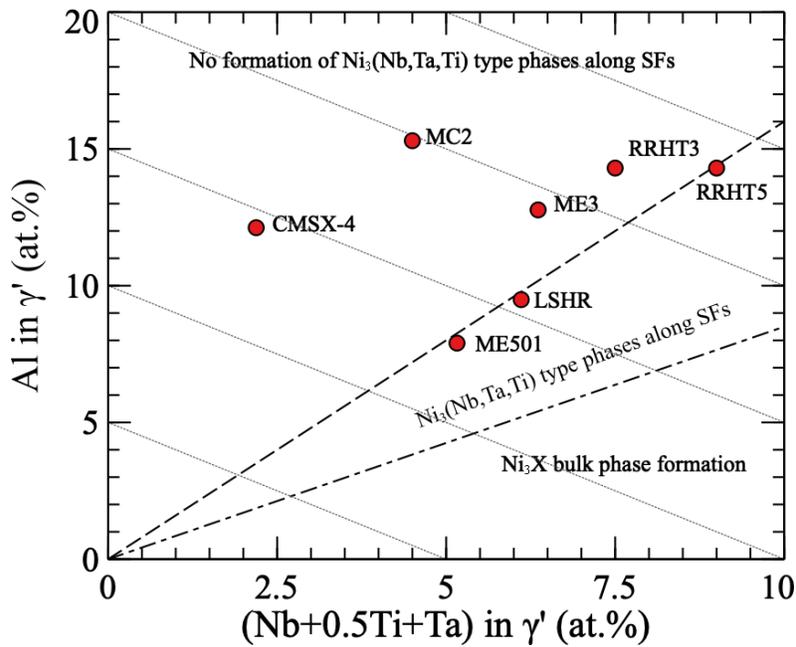

*Figure 14: Al concentration as a function of (Nb+0.5Ti+Ta) concentration for the alloys of the present study, as well as CMSX4 [25], MC2 [22], ME3 [21,48], LSHR [21], ME501[25] from the literature, showing that regions appear for alloys that form ordered phases along the stacking faults, suggesting a new design parameter.*

5. Conclusions

A polycrystalline Ni-based superalloy with two different Nb levels, namely RRHT3 (low-Nb) and RRHT5 (high-Nb), were studied. Their creep properties at 750°C under a 600 MPa load were tested, and the deformation mechanisms were investigated from a solute partitioning side, to better apprehend the critical consequences of chemistry variations on the mechanical properties. The conclusions can be summarized as follows:

- The RRHT5 alloy exhibits a desirable lower creep strain rate compared to RRHT3 alloy at 750°C under a 600 MPa load.
- The difference in creep performance is ascribed to the different types of crystal defects and their partitioning behavior. ECCI analyses confirmed the presence of

- stacking faults in the γ' precipitates of both crept alloys. Detrimental microtwins were observed only in RRHT3, promoting higher creep strain rate.
- APT analyses have shown that Co, Cr and Mo partition at dislocations for both alloys while Nb is depleted.
- Based on the present results coupled with literature survey, partitioning of Cr, Mo, Nb and Co at SISF and Nb and Co only at SESF was measured.
- Diffusion of solutes at the stacking faults, occurs in-plane with the fault as shown by APT analysis.
- With a local composition at the SESF up to 14 and 11.5 at.% for Co and Nb respectively and up to 17.5, 10, 2.8 and 0.3 at.% for Co, Nb, Cr and Mo at the SISF, it is likely that local phase transformation into the η-phase at the SESF and the χ-phase on the SISF occurs.
- The lower creep strain rate in the RRHT5 (high-Nb) alloy is hence rationalized by the combined presence of η-phase and χ-phase along the faults, hindering shearing of the γ' precipitates, and suppression of microtwinning.
- Using the results presented here and alloys of the literature, a design strategy is proposed, based on the (Nb+0.5Ti+Ta) composition. It is suggested that when this composition exceeds ~0.625 times the Al content, phase transformation along stacking faults is favored.

The present findings highlight that a compositional difference as small as the one between the two studied alloy, of 0.9 at.% Nb and 1.8 at.% Al, leads to drastic differences in the creep properties, that can be rationalized by the formation of ordered phases at the faults in one case, when it does not happen in the other. Future studies of various alloys, displaying or not phase transformation on SESF, will help refining the threshold of the proposed parameter, considering the (Nb+0.5Ti+Ta) composition, which could potentially be used as a criterion for the design of future superalloys with enhanced creep resistance.

**Acknowledgements**

Uwe Tezins & Andreas Sturm for their support to the FIB & APT facilities at MPIE. We are grateful for the financial support from the Max-Planck Gesellschaft via the Laplace project for both equipment and personnel (P.K. and B.G). P.K. also acknowledges financial support from the DFG SFB TR 103 through project A4. BG acknowledges financial support from the ERC-CoG-SHINE-771602. We acknowledge provision of the materials used in this study from Rolls-Royce Corporation.